\documentclass[aps,prd,notitlepage,showpacs,nofootinbib,superscriptaddress,twocolumn
,nofloats]{revtex4-1}

\usepackage{graphicx}
\usepackage[utf8]{inputenc} 
\usepackage{amsmath}
\usepackage{amssymb}
\usepackage{float}
\usepackage{comment}
\usepackage{slashed}
\usepackage[normalem]{ulem}

\usepackage[T1]{fontenc} 
\usepackage{parskip}
\usepackage{bm}         
\usepackage{array}
\usepackage{multirow}   
\usepackage{booktabs}   

\usepackage{textcomp}
\usepackage{amssymb}

\usepackage{enumerate}
\usepackage{caption}
\usepackage{subcaption}
\usepackage{ragged2e}
 \DeclareCaptionJustification{justified}{\justifying}
\usepackage{amsmath}
\usepackage{amssymb}

\usepackage[usenames,dvipsnames]{xcolor}
\usepackage[colorinlistoftodos]{todonotes}
\usepackage[colorlinks=true,citecolor=darkred,urlcolor=darkred, pdfborder={0 0 0}]{hyperref}
\usepackage[normalem]{ulem}

\definecolor{darkred}{rgb}{0.6,0,0}
\usepackage[colorinlistoftodos]{todonotes}

\definecolor{linkcolor}{rgb}{0,0,0.5}
 
\newcommand {\ignore}[1]{}

\definecolor{bostonuniversityred}{rgb}{0.8, 0.0, 0.0}

\def\gsim{\raise0.3ex\hbox{$\;>$\kern-0.75em\raise-1.1ex\hbox{$\sim\;$}}}
\def\lsim{\raise0.3ex\hbox{$\;<$\kern-0.75em\raise-1.1ex\hbox{$\sim\;$}}}

\def\SM{$\mathrm{SU(3)_c \otimes SU(2)_L \otimes U(1)_Y}$ }
\newcommand{\sm}{{Standard Model }}

\usepackage{soul}

\definecolor{mightnightblue}{RGB}{25,25,112}

\definecolor{brown}{rgb}{0.59, 0.29, 0.0}

\def\vev#1{\left\langle #1\right\rangle}
\def\SM{$\mathrm{SU(3)_c \otimes SU(2)_L \otimes U(1)_Y}$ }
\def\21{$\mathrm{SU(2)_L \otimes U(1)_Y}$}

\def\sm{standard model }

\newcommand{\AddrAHEP}{%
  AHEP Group, Institut de F\'{i}sica Corpuscular --
  C.S.I.C./Universitat de Val\`{e}ncia, Parc Cient\'ific de Paterna.\\
 C/ Catedr\'atico Jos\'e Beltr\'an, 2 E-46980 Paterna (Valencia) - SPAIN}

\newcommand{\AddrUNAM}{ {\it Instituto de F\'{\i}sica, Universidad Nacional Aut\'onoma de M\'exico, A.P. 20-364, Ciudad de M\'exico 01000, M\'exico.}}

\newcommand{\Addper}{{\it Perimeter  Institute  for  Theoretical  Physics,  Waterloo,  Ontario  N2L  2Y5,  Canada.}}

\begin{document}

\title{\boldmath\color{BrickRed} Dirac neutrinos from Peccei-Quinn symmetry: a fresh look at the axion}

\author{Eduardo Peinado}\email{epeinado@fisica.unam.mx}
\affiliation{\AddrUNAM}
\author{Mario Reig}\email{mario.reig@ific.uv.es}
\affiliation{\Addper}
\affiliation{\AddrAHEP}
\author{Rahul Srivastava}\email{rahulsri@ific.uv.es}
\affiliation{\AddrAHEP}
\author{Jose W. F. Valle}\email{valle@ific.uv.es}
\affiliation{\AddrAHEP}
\vspace{0.7cm}
\date{\today}

\begin{abstract}
  \vspace{0.3cm}
We show that a very simple solution to the strong CP problem naturally leads to Dirac neutrinos.
Small effective neutrino masses emerge from a type I Dirac seesaw mechanism.
Neutrino mass limits probe the axion parameters in regions currently inaccessible to conventional searches.
\end{abstract}

\keywords{Peccei-Quinn symmetry, Axion, Neutrinos}
\maketitle



\section{Introduction}


Despite its tremendous success, the \sm has many drawbacks and theoretical loose ends.
Amongst them the lack of neutrino masses and mixings~\cite{Valle:2015pba}, and of a viable dark matter candidate~\cite{Bertone:2004pz}. 
Both issues require new physics, beyond the standard model.
Moreover, the \sm also leaves unexplained the lack of CP violation in the strong interaction~\cite{Peccei:1977hh,Weinberg:1977ma,Wilczek:1977pj}.
Aware that the list of shortcomings is much longer, here we focus on whether the above three aspects may be closely interconnected in the context of the Peccei-Quinn (PQ) mechanism. 
In fact, there have already been recent attempts to connect it to neutrino mass generation, both in the Majorana \cite{Dasgupta:2013cwa,Bertolini:2014aia,Reig:2018yfd,Suematsu:2017kcu} and Dirac \cite{Chen:2012baa,Gu:2016hxh,Carvajal:2018ohk} frameworks.\\[-.2cm]

In axion models quarks carry a non-zero PQ charge, so two Higgs doublets $H_u$ and $H_d$ are typically required.
In order to be phenomenologically viable, the PQ symmetry must break at high energies, implying the need for a \SM singlet scalar boson, carrying PQ charge, denoted as $\sigma\sim (1,1,0)$.
Moreover, the $H_{u,d}$ fields must couple to $\sigma$ in such a way that the only U(1) symmetries are $\mathrm{U(1)_Y\otimes U(1)_{PQ}}$.
There are two ways of doing this, depending on the form of the mixing terms in the scalar potential. 
These yield two possibile choices for the PQ charge of $\sigma$, that may be taken as 2 or 4, for one of the simplest Higgs doublet charge assignments. 
When the PQ charge is 2, the spontaneous breaking of the Peccei-Quinn symmetry can be connected to the breaking of lepton number by two units~\cite{Mohapatra:1982tc} leading to Majorana neutrinos \cite{Schechter:1980gr}. \\[-.2cm]

In this letter we challenge the view that linking spontaneous Peccei-Quinn symmetry breaking to neutrino mass generation leads to Majorana neutrinos.
This is achieved by making the alternative choice for the PQ charge. We show how it leads to a novel class of minimal axion models that effectively imply Dirac neutrinos.
For definiteness, we take as reference the simplest DFSZ axion scheme~\cite{Dine:1981rt,Zhitnitsky:1980tq}, taking the associated field with PQ charge $4$.
We also provide the simplest UV-completion of the new ``Diraxion'' scheme, where the neutrino masses are naturally small, implemented through the type-I Dirac seesaw mechanism.
  Neutrino mass limits, such as the recent one of the Katrin tritium $\beta$ decay experiment~\cite{Aker:2019uuj}, provide new ways to probe the axion parameter space.


\section{Minimum setup} 
\label{sec:minimum-setup} 


As mentioned, here we depart from the canonical choice for axion quantum numbers.
We focus on the minimal DFSZ model where the Higgs fields $H_u$ and $H_d$ have PQ charge 2, while the symmetry breaking field $\sigma$, associated with the axion field, has PQ charge $4$.
In this case there is a term in the potential of the form 
\begin{equation}
 V_{mix}(H_u,H_d,\sigma) \propto H_u H_d \sigma^*   \,,  
\end{equation}
involving a dimensionful coupling. The crucial observation is that with such assignment as in Table~\ref{tab1} there is no way to form the dimension-five Weinberg operator for the light neutrino masses, nor any other operator with powers of $\sigma$ and or powers of $H_{u}$ and $H_{d}$. 
Indeed, we first notice that with two Higgs doublets there are three different dimension-five Weinberg operators~\footnote{Barring multiple Higgses, the uu and dd-type contractions vanish identically, however this does not change our argument.}, of the form 
\begin{equation}\begin{array}{lcr}
&{\mbox{Operator}}&\mbox{PQ charge}\\\\
{\mathcal L}_{dim~5}\sim& 
\left\{\begin{array}{l}
\frac{LL\tilde{H_u} \tilde{H_u} }{\Lambda_{UV}}\\\\
\frac{LL\tilde{H_u} H_d }{\Lambda_{UV}}\\\\
\frac{LLH_d H_d }{\Lambda_{UV}}
\end{array}\right.&\begin{array}{r}1+1+(-4)={-2}\\\\
1+1+(0)={+2}\\\\1+1+(4) = +6
\end{array}\end{array}\label{dim5}
\end{equation}
Clearly, none of these operators is invariant under the PQ symmetry. 
As a result, from Eq. \eqref{dim5}, there is no way to construct an operator invariant under $\mathrm{U(1)_{PQ}}$ and \sm symmetries simultaneously.
One can verify that this argument also extends to all the higher order effective operators that could potentially generate Majorana masses for neutrinos.
To see this consider all possible gauge invariant contractions of the scalars and their PQ charges, which are shown in parenthesis in Eq.~\eqref{eq:pq-charge},
\begin{equation}\begin{array}{lrlr}
\sigma^n  & (4 n); & \quad
(\sigma^*)^n & (-4 n);\\
(H_u H_d)^n & (4 n);  & \quad
(H_u H_d)^{*n} & (-4 n);\\
(H_u^\dagger H_u)^n & (0); & \quad
(H_d^\dagger H_d)^n & (0).\\\end{array}
\label{eq:pq-charge}
\end{equation}
In order to form a neutrino Majorana mass term, one must insert at least one of the scalar contractions shown in Eq.~\eqref{eq:pq-charge} into Eq.~\eqref{dim5}.
Since all these gauge invariant combinations carry either zero or $4n$ PQ charge ($n = 1,2,3\dots$), no combination can lead to a PQ invariant Majorana mass generating operator.   
Hence the only option is for neutrinos to be Dirac particles.\\[-.2cm]

We now turn to the question of generating finite Dirac neutrino masses. 
One option is to include the ``right-handed'' neutrinos $\nu_{Ri}$ with PQ charge 1. In this case the Yukawa Lagrangian is
\begin{eqnarray}\label{Lag1}
 {\cal L}_{Y} & = & y_{ij}^{u}\bar{Q}_{i}H_{u}u_{j} + y_{ij}^{u} \bar{Q}_{i}H_{d} d_{j} + y_{ij}^{l}\bar{L}_{i}H_{d}l_{j}  \nonumber \\
  & + &    y_{ij}^{\nu}\bar{L}_{i}H_{u} \nu_{Rj} + h.c.~, 
\end{eqnarray}
so that neutrinos are Dirac particles and the Yukawa couplings $y_{{ij}}^{\nu}$ must be of the order ${\cal O}(10^{-12})$ in order to account for the recent Katrin bound.
Such a small coupling suggests the need for a dynamical explanation.
Let us now explore the possibilities to generate such a small coupling in a natural way.

\begin{widetext}
 \begin{center}
\begin{table}[t]
\begin{tabular}{|c | c c c   c c| c c c|}
  \hline  
\hspace{0.4cm} Fields/Symmetry  \hspace{0.4cm} & \hspace{0.4cm} $Q_i $ \hspace{0.4cm}  & \hspace{0.4cm} $ u^c_i$ \hspace{0.4cm}  & \hspace{0.4cm} $d^c_i$ \hspace{0.4cm}   & \hspace{0.4cm}  $L_i$  \hspace{0.4cm}  & \hspace{0.4cm} $l^c_i$ \hspace{0.4cm} &  \hspace{0.4cm} $H_u$ \hspace{0.4cm} & \hspace{0.4cm} $H_d$ \hspace{0.4cm} & \hspace{0.4cm} $\sigma$ \hspace{0.4cm} \\
\hline
$SU(2)_L\times U(1)_Y$   &  (2,1/6)  &  (1,-2/3)  &  (1,1/3)  &  (2,-1/2)  &  (1,1)    &        (2,-1/2)                &      (2,1/2)               &          (0,0)           \\   
\hline
$U(1)_{PQ}$    &  1     &   1     &  1     &  1     &  1    &         2               &     2              &         4             \\
\hline
  \end{tabular}
\caption{Quantum numbers in the DFSZ axion model.  All fermions are left-chiral.}
 \label{tab1} 
\end{table}
\end{center}
\end{widetext}


\section{Type I Dirac seesaw}
\label{sec:type-i-dirac}


In the presence of adequate protective symmetries, there are many pathways to generate naturally small Dirac neutrino masses.
This can be done \textit{\`a la seesaw}, using dimension-5 and/or dimension-6 operators~\cite{Ma:2016mwh,CentellesChulia:2018gwr,CentellesChulia:2018bkz}.
Many full-fledged UV-complete seesaw-based as well as radiative theories of Dirac neutrino mass generation have been proposed~\cite{Ma:2014qra,Chulia:2016ngi,Bonilla:2017ekt,CentellesChulia:2017koy,Bonilla:2016diq,Reig:2016ewy,Bonilla:2018ynb}.
The task here is to do the same using realizations of the PQ symmetry. For definiteness we stick to the type I seesaw mechanism. 

\begin{figure}[h]
    \centering
\includegraphics[scale=.60]{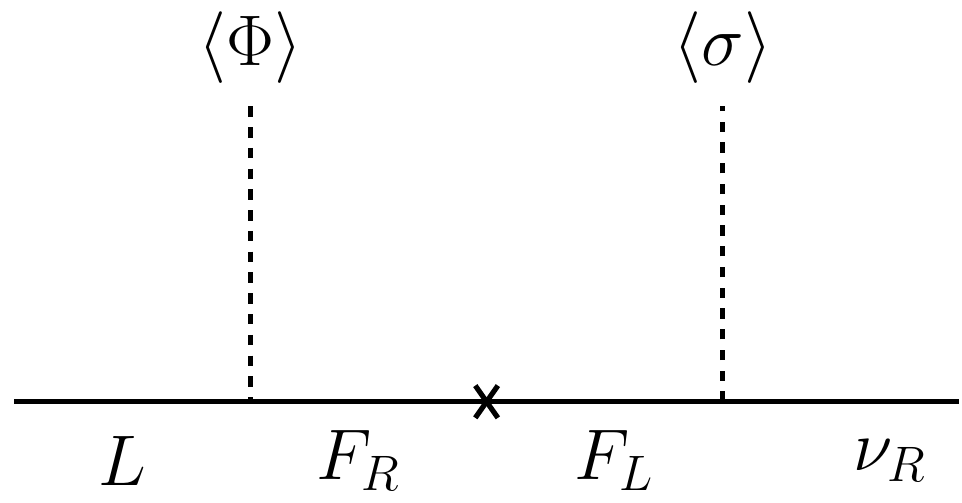}
 \caption{ Dirac neutrino mass generation diagram.}
    \label{fig:TID}
  \end{figure}

In order to generate Dirac neutrino masses \textit{\`a la seesaw}, the last term in Eq.~\eqref{Lag1} must be forbidden.
This can happen in a simple and natural way if the PQ charge of $\nu_{Ri}$ differs from that of the lepton doublets $L_i$.
For definiteness consider now the case in which the ``right-handed'' neutrinos have charge $-5$ under the $\mathrm{U(1)_{PQ}}$ symmetry.
In such a case the tree-level Dirac mass term is PQ--forbidden, but a dimension 5 operator is allowed
\begin{equation}\label{Dim5Op}
 {\cal L}_{{dim5}}^{D} =  y_{ij}^{\nu}\bar{L}_{i}H_{u} \nu_{Rj}\frac{\sigma}{\Lambda_{UV}}  + h.c.  
\end{equation}
A very simple way to UV-complete this operator is through the Type I Dirac seesaw mechanism.
To this end we simply extend the model by including three sequential pairs of chiral fermions $F_{L i}$ and $F_{R i}$, as shown in Table (\ref{tab2}).

The relevant neutrino mass Lagrangian will be 
\begin{equation}\label{TI}
{\cal L}_{typeI} = \lambda_{ij}\bar{L}_{i} H_{u} F_{Rj} +  \kappa_{{ij}} \bar{F}_{Li} 
\sigma \nu_{Rj}  + M_{ij} \bar{F}_{Li} F_{Rj} + h.c  
\end{equation}

Notice that, unlike the Majorana case where one typically has
\begin{equation}
m^{\rm{Majorana}}_\nu\sim v_{EW}^2/f_a\,, 
\end{equation}
for the Dirac case one obtains, from Eq. \eqref{Dim5Op}~\footnote{This resembles reference~\cite{Mohapatra:1986ks}, but now it is the PQ symmetry that enforces the Diracness of the neutrinos.}:
\begin{equation} \label{neutrino_mass}
 m^{\rm{Dirac}}_{\nu}\sim v_{EW} f_a/\Lambda_{UV}\,. 
\end{equation}

  For axion decay constants around $f_a\sim 10^8$ GeV, the minimum value allowed from astrophysicsl constraints~\cite{Irastorza:2018dyq}, and reasonable Yukawa couplings $\sim 10^{-3}$, neutrino masses in the eV scale would correspond to the UV scale $\Lambda_{UV} \sim M_{GUT}$, which is suggestive. 

One sees that, for fixed Yukawa couplings, our axion prefers smaller values of the breaking scale $f_a$, corresponding to higher (smaller) axion (neutrino) masses. 
\begin{widetext}

\begin{center}
\begin{table}[t]
\begin{tabular}{|c | c c c  c c c|cc| c c c|}
  \hline
\hspace{0.3cm} Fields/Symmetry \hspace{0.3cm} & \hspace{0.3cm} $Q_i $ \hspace{0.3cm}  & \hspace{0.3cm} $ u^c_i$ \hspace{0.3cm}  & \hspace{0.3cm} $d^c_i$ \hspace{0.3cm} & \hspace{0.3cm} $L_i$ \hspace{0.3cm} & \hspace{0.3cm} $l^c_i$ \hspace{0.3cm} & \hspace{0.3cm} $\nu^c_{i}$ \hspace{0.3cm} & \hspace{0.3cm} $ F_{i}$ \hspace{0.3cm} & \hspace{0.3cm} $F^c_{i}$ \hspace{0.3cm} & \hspace{0.3cm} $H_u$ \hspace{0.3cm}  & \hspace{0.3cm} $H_d$ \hspace{0.3cm}   & \hspace{0.3cm} $\sigma$  \hspace{0.3cm}  \\
\hline
$SU(2)_L\times U(1)_Y$ &  (2,1/6)  &   (1,-2/3)  &  (1,1/3)   &  (2,-1/2)  &  (1,1)   &  (1,0)  &  (1,0)  &  (1,0)  &        (2,-1/2)                &      (2,1/2)               &          (0,0)           \\   
\hline
$U(1)_{PQ}$ &  1  &   1  &  1  &  1  &  1  &   5  &  -1  &  1  &         2               &     2              &         4             \\
\hline
  \end{tabular}
  \caption{ Proposed assignment of our ``Diraxion'' model. The sequential chiral fermions $F$ and $F^c$ merge to form the Dirac fermions that mediate neutrino mass generation through
    Fig.~\ref{fig:TID}. } 
 \label{tab2} 
\end{table}\end{center}
\end{widetext}


\section{Shrinking the axion landscape} 
\label{sec:shrink-axion-landsc}


The axion coupling to photons is determined by the QCD anomaly and is given as \cite{Kaplan:1985dv,Srednicki:1985xd,diCortona:2015ldu} 
\begin{equation}\label{axion-photon coupling}
g_{a\gamma}=\frac{\alpha_{EM}}{2\pi f_a}\left(E/N-1.92\right)\,.
\end{equation}

For reasonable values of the $E/N$ ratio (which in our model is 8/3) this coupling can be probed in a variety of experiments~\cite{Irastorza:2018dyq}. 

A characteristic feature of our scenario is that neutrino mass experiments also play a role.
To see this, the first thing to note is that the messenger mass $M_\chi\sim\Lambda_{UV}$ can be substantially higher than that in the usual type-I Majorana seesaw mechanism. 
Indeed, it can lie at the Planck scale~\footnote{For Planck-scale lepton number violation schemes see~\cite{deGouvea:2000jp,Ibarra:2018dib}.} $\Lambda_{UV}\sim M_{Planck}$ or be associated to some Grand unification (GUT) group, $\Lambda_{UV}\sim M_{GUT}$.

The unusual dependence in Eq.~\eqref{neutrino_mass}~\footnote{Notice that this dependence is quite generic, and applies to any UV completion of the effective operator in Eq. \eqref{Dim5Op}.}, with the neutrino mass, scaling as $m_\nu\propto f_a$ ($f_a$ the axion decay constant), implies that the heavier -- and hence the more \textit{strongly} coupled -- is the QCD axion, the lighter are the neutrinos.
This offers the possibility of probing the QCD axion physics with neutrino physics considerations. 

\begin{widetext}

\begin{figure}[h!]
	\centering
	\includegraphics[width=0.8\textwidth]{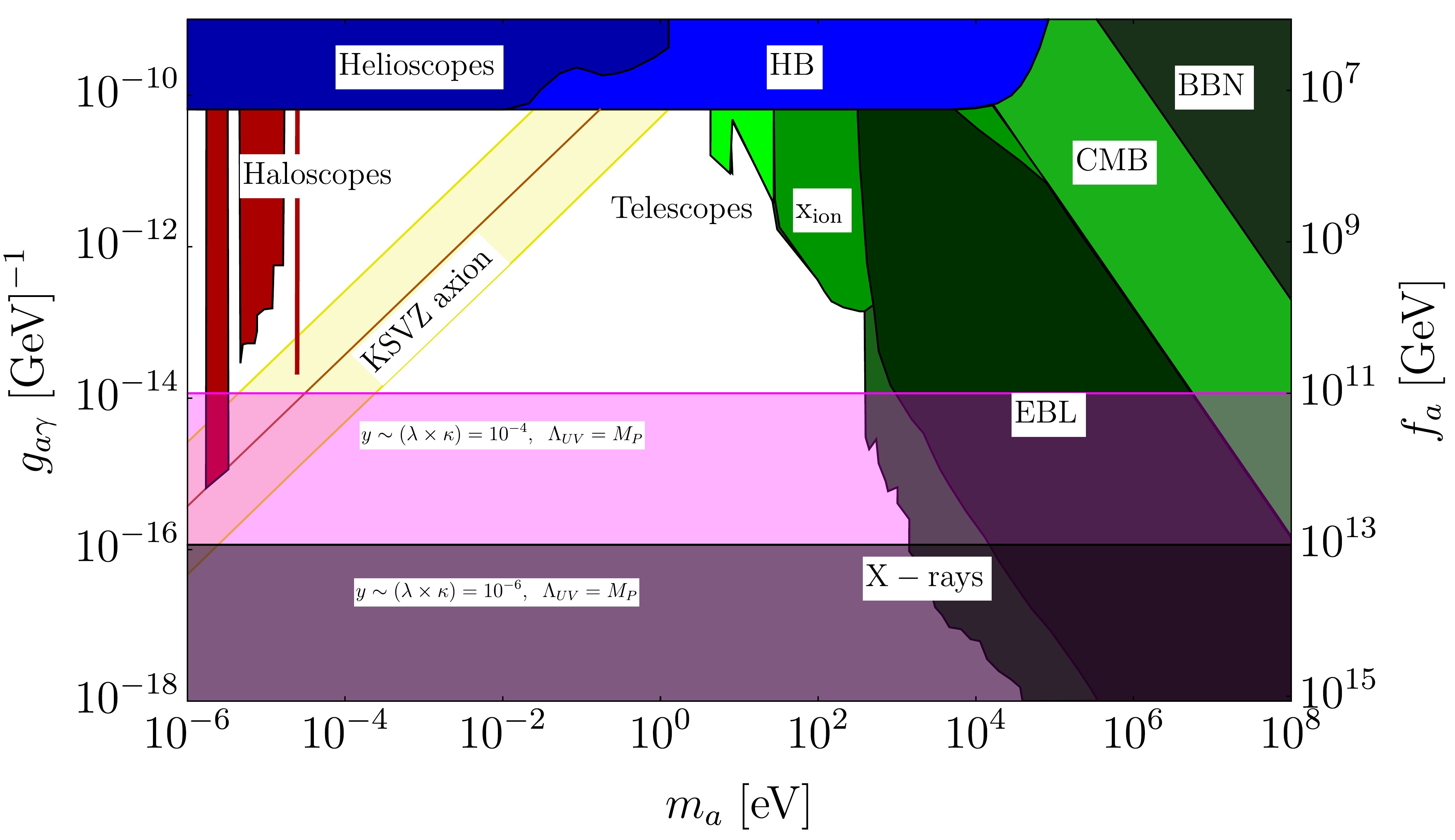}
	\caption{Landscape of axion parameters ($m_a$, $g_{a\gamma}$), adapted from \cite{Irastorza:2018dyq}. 
 We show the constraints from cosmology \cite{Cadamuro:2010cz} (green), astrophysics \cite{Anastassopoulos:2017ftl,Ayala:2014pea} (blue) and haloscopes \cite{Asztalos:2009yp6,Brubaker:2016ktl} (dark red), together with the predicted QCD band (shaded yellow) and the KSVZ line (brown).
 The pink horizontal band illustrates neutrino masses obtained by setting $\Lambda_{UV}$ to the Planck scale $M_{P}$ and fixing Yukawa couplings in the indicated range. 
One sees that the Katrin limit $m_\nu \leq 1.1$ eV \cite{Aker:2019uuj} would place an upper bound on the axion decay constant, hence a lower bound on the axion-photon coupling.  
}\label{fig:axion_plot} 
          \end{figure}
          
\end{widetext}

We start by setting $\Lambda_{UV}$ to the Planck scale $M_{P}$, and fixing Yukawa couplings (see Eq. \eqref{TI}) within a reasonable range $10^{-2}-10^{-3}$, such that the product $y\sim\lambda \times \kappa$ (see Eq. \eqref{Dim5Op} and Eq. \eqref{TI}) lies between $10^{-4}$ and $10^{-6}$.
Once the Yukawa couplings and the UV scale $\Lambda_{UV}$ are fixed, one can use neutrino mass bounds, such as the recent Katrin results~\cite{Aker:2019uuj} to probe part of the axion parameters, as seen in Fig.~\ref{fig:axion_plot}.
Indeed, the new neutrino mass upper bound would place an upper bound to $f_a$, and hence a lower bound to the axion-photon coupling due to Eq. \eqref{axion-photon coupling}.
One sees that neutrino experiments such as Katrin can probe the parameter space towards the bottom, where no experiment can look for axions directly.  
Likewise, this sentivity would also apply for generic ALPs coupled to photons, as indicated in Fig.~\ref{fig:axion_plot}.
In short, our model illustrates the interplay and potential complementarity between axion searches and neutrino experiments, which is a characteristic feature of our proposal.

In contrast, note that by choosing a lower $\Lambda_{UV}=M_{GUT}$, the allowed parameter space in Fig.~\ref{fig:axion_plot} would be substantially reduced.

\section{Unification}
\label{sec:unification}


We now comment on the fact that the quantum numbers in Tables~(\ref{tab1}) and (\ref{tab2}) are suggestive of the idea of unification. Here we sketch an SO(10) embedding.
By appropriate assignment of the PQ charges one can achieve the seesaw mechanism in Fig.\ref{fig:TID} within the framework of SO(10). 
For example, in addition to the three \sm families embedded in the $\textbf{16}^F$ spinors, one introduces the fields: 
\begin{equation}
 \textbf{1}^S_4\,,\,\,\textbf{10}^S_2\,,\,\,\textbf{16}^S_0\,,\,\,\textbf{1}^F_{-1}\,,\,\,\textbf{1}^F_{5}\,,\,\,
\nonumber
\end{equation}
where $F,S$ stand for fermions and scalars, respectively.
\begin{figure}[h!]
\includegraphics[scale=.6]{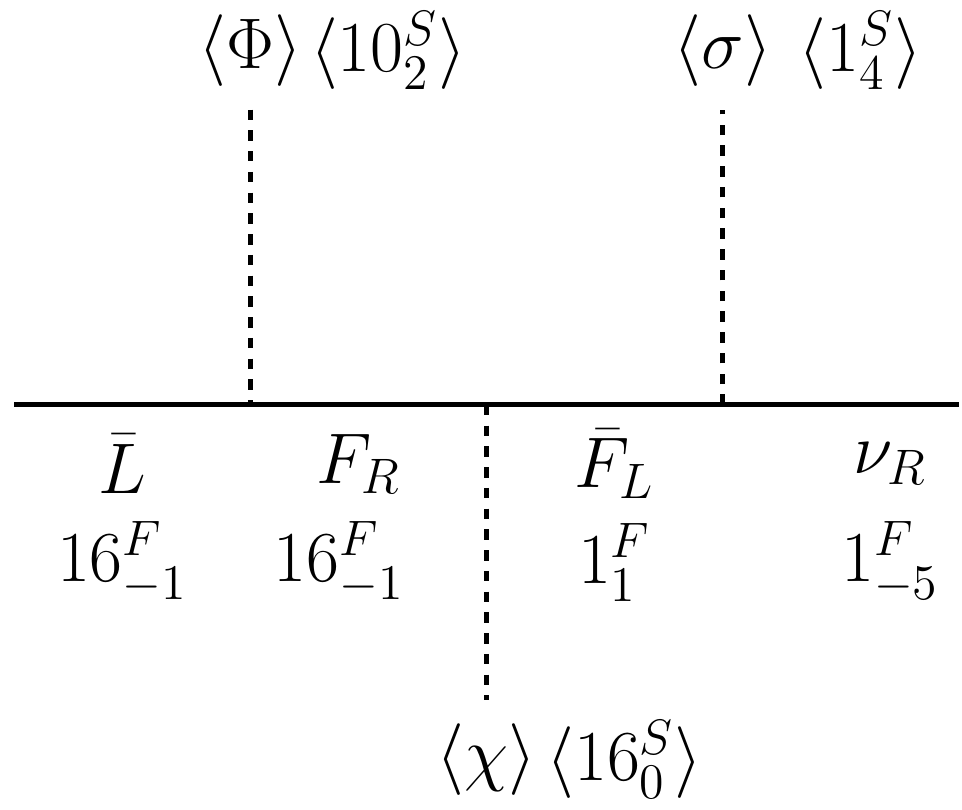}
 \caption{Dirac neutrino mass generation diagram in SO(10), indicating also the PQ charges.}
    \label{fig:TIDg}
\end{figure}

Notice the unusual embedding of the ``right-handed'' neutrino: while the $F_{R}$ field lies in the $\mathbf{16^F}$ spinor, the ``right-handed'' piece ($\nu_{Ri}$) of the \sm neutrino comes as an SO(10) singlet, with PQ charge -5, $\textbf{1}^F_{-5}$. 
The scalar spinor $\textbf{16}^S_0$ is responsible for giving a Dirac mass to $F$, breaking SO(10) down to SU(5), but preserving the PQ symmetry.
The latter will be broken by $\textbf{1}^S_4$. As usual, the $H_u$, $H_d$ fields will reside inside the $\textbf{10}^S_{2}$.
They will give masses to the charged fermions and contribute to Dirac neutrino mass generation,  which in the Dirac basis of $(\bar{\nu}_L , \bar{F}_L)^T$ and 
$(\nu_R , F_R )$ can be written as
\begin{eqnarray}
 M_{seesaw} & = & \left(\begin{matrix}
0 &  \lambda \vev{ \Phi }_{10^S_2}\\
\kappa \vev{\sigma}_{1^S_4}  & \kappa_F \vev {\chi}_{16^S_0} \\
\end{matrix}\right)\, ,
 \label{eq:mass-matrix}
\end{eqnarray}
where $\lambda, \kappa, \kappa_F$ are the Yukawa couplings.
The resulting light seesaw neutrino mass is similar to Eq.~\eqref{neutrino_mass} above, by identifying  $\Lambda_{UV}$ in Eq.~\eqref{Dim5Op} 
with the scale of the breaking SO(10) $\to$ SU(5), $\Lambda_{UV} \sim M_{GUT}$. 
Of course, for a fully consistent GUT construction additional multiplets would be needed, though are not expected to affect the mass mechanism proposed above.

\section{Conclusions}

Here we have proposed a simple solution to the strong CP problem in which neutrinos are naturally predicted to be Dirac fermions.
Small effective neutrino masses, directly proportional to the PQ breaking scale, emerge from a type I Dirac seesaw mechanism.
By embedding the DFSZ axion into a full-fledged neutrino mass generation setup, we have shown how neutrino mass limits provide independent probes of the axion parameters.
Interestingly enough, due to the special form of Eq.~\eqref{neutrino_mass}, this new probe is complementary to those accessible to the existing axion searches.
Our ``Diraxion'' scheme is compatible with the idea of unification, providing a counter-example to the common belief that GUTs, and in particular SO(10),
must lead to Majorana neutrinos. 

\begin{acknowledgments}
	
Work supported by the Spanish grants SEV-2014-0398 and FPA2017-85216-P (AEI/FEDER, UE), PROMETEO/2018/165 (Generalitat Valenciana) and the Spanish Red Consolider MultiDark FPA2017-90566-REDC. This work is also supported by the Mexican grants DGAPA-PAPIIT IN107118 (M\'exico) and CONACyT CB-2017-2018/A1-S-13051 (M\'exico). M.R. is grateful for the hospitality of the Instituto de F\'isica at UNAM (Mexico) and the Visiting Graduate Fellowship program at Perimeter Institute where part of this work was carried out. Research at Perimeter Institute is supported in part by the Government of Canada through the Department of Innovation, Science and Economic Development Canada and by the Province of Ontario through the Ministry of Economic Development, Job Creation and Trade.  M.R. thanks Luca Visinelli for useful comments.
\end{acknowledgments}

\bibliographystyle{utphys}
\bibliography{bibliography}
\end{document}